\documentclass[10pt,journal,compsoc]{IEEEtran}

\ifCLASSOPTIONcompsoc
  \usepackage[nocompress]{cite}
\else
  \usepackage{cite}
\fi

\usepackage{amsmath}
\usepackage{algorithmic}
\usepackage{url}
\usepackage{hyperref}
\usepackage{graphicx}
\usepackage{amsfonts}
 \usepackage[table,xcdraw]{xcolor}
\usepackage{multirow}
\usepackage{microtype}

\usepackage[ruled,linesnumbered]{algorithm2e}

\SetAlFnt{\small}
\SetAlCapFnt{\small}
\SetAlCapNameFnt{\small}
\SetAlCapHSkip{0pt}

\begin{document}
%
\title{\huge
Deterministic Linear Time Constrained\\
Triangulation using Simplified Earcut}

%
%
%
%

\author{Marco~Livesu, Gianmarco~Cherchi, Riccardo~Scateni, Marco~Attene}

\markboth{Journal of \LaTeX\ Class Files,~Vol.~14, No.~8, August~2015}%
{Shell \MakeLowercase{\textit{et al.}}: Bare Demo of IEEEtran.cls for Computer Society Journals}

\IEEEtitleabstractindextext{%
\begin{abstract}
Triangulation algorithms that conform to a set of non-intersecting input segments typically proceed in an incremental fashion, by inserting points first, and then segments. Inserting a segment amounts to: (1) deleting all the triangles it intersects; (2) filling the so generated hole with two polygons that have the wanted segment as shared edge; (3) triangulate each polygon separately.
In this paper we prove that these polygons are such that all their convex vertices but two can be used to form triangles in an earcut fashion, without the need to check whether other polygon points are located within each ear.  The fact that any simple polygon contains at least three convex vertices guarantees the existence of a valid ear to cut, ensuring convergence.
Not only this translates to an optimal deterministic linear time triangulation algorithm, but such algorithm is also trivial to implement. We formally prove the correctness of our approach, also validating it in practical applications and comparing it with prior art.
\end{abstract}

\begin{IEEEkeywords}
constrained triangulation, tessellation, segment insertion, earcut, CDT
\end{IEEEkeywords}}

\maketitle
\IEEEdisplaynontitleabstractindextext
\IEEEpeerreviewmaketitle


\newcommand{\cino} [1]{{\color{magenta}	Cino: #1}}
\newcommand{\giammi} [1] {{\color{red} Giammi: #1}}
\newcommand{\jaiko} [1]{{\color{blue}	Jaiko: #1}}
\newcommand{\com} [1]{{}} 
\newcommand{\edit} [1]{{\color{black}			#1}} 

\newcommand{\rev} [1]{{\color{blue}			#1}} 

\newcommand*{\qed}{\null\nobreak\hfill\ensuremath{\square}}%

\IEEEraisesectionheading{\section{Introduction}}
\label{sec:introduction}

\begin{figure*}
	\centering
	\includegraphics[width=\linewidth]{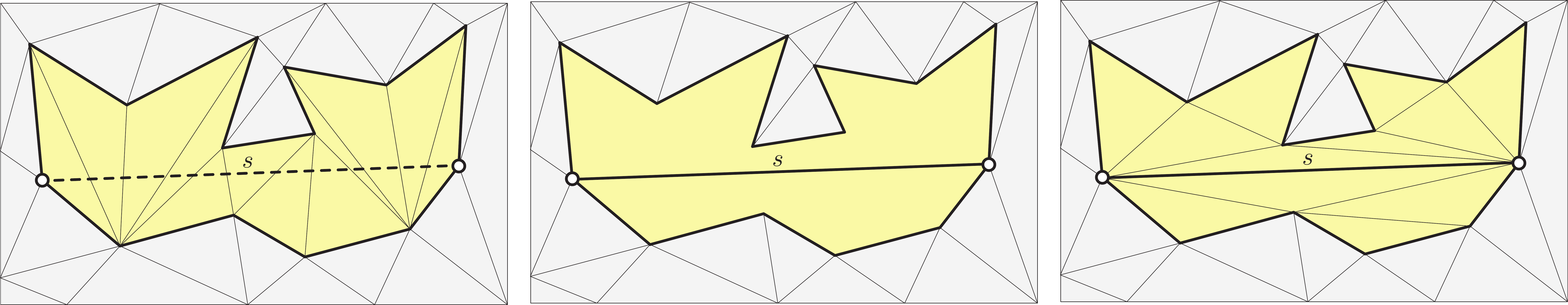}
	\caption{Inserting a constrained segment $s$ in a previously existing triangulation requires removing all the triangles intersected by it (in yellow), revealing two polygonal pockets having $s$ as base. These pockets may be non convex, but endow an important property: a portion of $s$ can be seen from any of their interior points. This can be proved by observing that these polygons are fully formed by portions of triangles that are cut by $s$, therefore the convexity of each sub-polygon guarantees visibility. In practice, this avoids the generation of curl-like concavities, a property that we exploit to speed up their re-triangulation.}
	\label{fig:pocket}
\end{figure*}

The generation of triangulations that conform to a given set of line segments is at the basis of many tools in scientific computing~\cite{shewchuk1996triangle,boissonnat2000triangulations}. A typical approach to the construction of a constrained triangulation consists in computing a generic triangulation of all the segment endpoints, and then incorporate the segments. Adding a segment connecting two vertices of a previously existing triangulation requires to perform two operations: (i) detect and remove all the triangles that are intersected by the segment; (ii) fill the so generated poygonal pocket, triangulating two sub-polygons that have the wanted segment as shared basis (Figure~\ref{fig:pocket}). In this short paper we focus our attention on this latter operation, sheding some new light on this classical computational geometry problem, and ultimately proposing a simple yet computationally optimal solution that has been surprisingly overlooked until now.

Our main intuition is that all the polygons that arise in the context of segment insertion belong to a restricted class of simple polygons which cannot contain severe (\textit{curl-like}) concavities. We exploit this property to devise a straightforward triangulation algorithm that proceeds in an earcut fashion~\cite{eberly2008triangulation}, forming triangles by cutting one ear at a time. We were able to prove that for any polygon in our class of interest all the convex vertices but two form valid ears which do not contain any other vertex inside, hence can be used to form triangles right away. We also prove that any such polygon contains at least three convex vertices, thus guaranteeing convergence. Putting all these ingredients together yields a triangulation algorithm which is a simplified version of the classical earcut, from which we omitted any point-in-triangle test. This simplification not only makes the algorithm even simpler to implement, but it also makes it run in deterministic linear time, on par with the best known triangulation algorithm~\cite{chazelle1991triangulating} which, conversely, is extremely difficult to implement.

Our linearized earcut method advances the state of the art in the field, which comprised either optimal algorithms that were complex to implement, or algorithms that were easier to implement (though still less easy than earcut) but had sub optimal asymptotic complexity (Section~\ref{sec:related}).

In Section~\ref{sec:earcut_quad} we describe the basis of the classical earcut algorithm, which has $O(n^2)$ complexity. In Section~\ref{sec:earcut_linear} we introduce our simplified version, demonstrating that it runs in deterministic linear time and also proving its correctness in Section~\ref{sec:correctness}. In Section~\ref{sec:results} we report on numerical tests we performed on our method, also comparing with the most recently published method in the field, proposed by Shewchuk and Brown in~\cite{shewchuk2015fast}.



\section{Prior works}
\label{sec:related}

Finding efficient methods to triangulate a polygon has been a foremost problem in computational geometry and computer graphics since decades. Before 1978 no efficient methods were known, and the only approach to triangulation was brute force. Brute force methods -- of which earcut~\cite{eberly2008triangulation} is a popular representative -- are very easy to implement, but at the same time they are inefficient, and can only achieve $O(n^2)$ complexity. 
The first attempt to efficiently triangulate a polygon occurred in 1978~\cite{garey1978triangulating}, and the proposed algorithm had $O(n \log n)$ time complexity. For a certain period it was thought that triangulation was a problem as hard as sorting, and no better algorithms could be devised. Asano et al.~\cite{asano1986polygon} showed that this bound is optimal for polygons with holes, but does not apply to simple polygons. Fournier and Montuno showed that the decomposition of a simple polygon into trapezoidal elements (\textit{trapezoidation}) is equivalent to triangulation, and that each trapezoid could be triangulated in $O(n)$~\cite{fournier1984triangulating}. At that time the best trapezoidation algorithm had $O(n \log n)$ complexity, which was therefore also a bound for triangulation. In the subsequent years various researchers focused their attention to trapezoidation as a mean to improve tiangulation, until in 1988 Tarjan and Van Wyk~\cite{tarjan1988n} showed that a trapezoidation (hence a triangulation) could be obtained in $O(n \log \log n)$.
In their paper, Tarjan and Van Wyk open about the possibility to achieve linear complexity in the near future, and in 1991 Chazelle proposed a deterministic linear time algorithm~\cite{chazelle1991triangulating} based on a very sophisticated trapezoidation technique. The cost these algorithms pay for their extreme efficiency is algorithmic complexity. Quoting~\cite{shewchuk2015fast}, Chazelle's work \textit{"is celebrated as a theoretical breakthrough, but is considered too complicated for practical use"}. Chazelle himself closed his famous article raising the question of whether there exist simpler algorithms that would allow to triangulate a polygon in optimal time. This question was partly answered in~\cite{amato2001randomized,amato2000linear}. However, these methods obtained only \textit{expected} linear time complexity using randomized approaches, but are still non optimal in the worst case.

General purpose algorithms have a very rich literature, and no major improvements have been registered in recent years. Our work does not aim to provide a contribution in this regard, but is rather linked to a parallel line of works, which focus on a specific application (constrained triangulation in our case). 
Restricting their applicability to a narrower class of inputs, these methods obtain efficiency with simpler algorithms that are easier to implement. 
Constrained triangulations are widely used in scientific computing, and a variety of methods tailored for them have been proposed over the years.
Anglada~\cite{anglada1997improved} extended the work of De Floriani and Puppo~\cite{de1992line}, proposing a simple $O(n \log n)$ algorithm for on-line segment insertion. According to~\cite{shewchuk2015fast} this is the easiest method to implement for this class of problems, but it has $O(n^2)$ complexity in the worst case. Methods that run in deterministic $O(n \log n)$ time are also available~\cite{kao1992incremental,lee1986generalized}. The state of the art in the field is~\cite{shewchuk2015fast}, which combines simplicity and efficiency, obtaining expected linear time complexity with a randomized approach. A method that runs in deterministic linear time also exists~\cite{chin1998finding}, but it is based on trapezoidation and is complicated to implement. We show that a trivial modification of a brute force method like earcut leads to optimal deterministic linear time complexity, and that our proposed modification even simplifies the original algorithm in terms of coding effort. To this end, not only our method has optimal complexity, but it is also easier to implement than any known technique, including brute force algorithms.

%
\begin{figure*}[t]
	\centering
	\includegraphics[width=\linewidth]{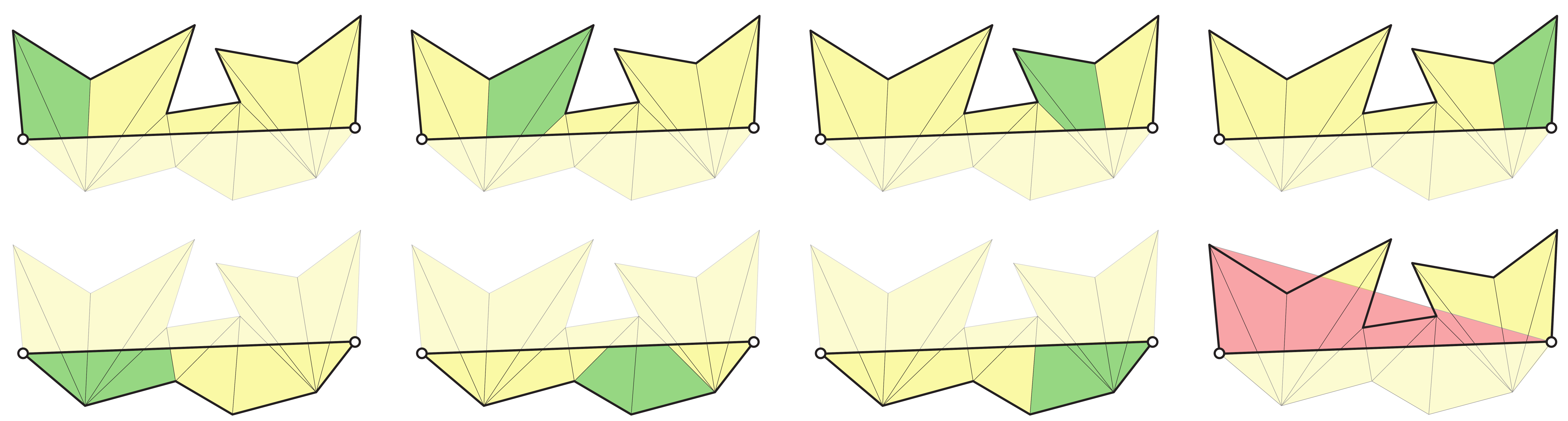}
	\caption{All internal convex vertices of the two polygons based upon the constrained segment (dashed line) define a convex sub-polygon (in green) that guarantees they are valid ears. Bottom right: the extrema of the constrained segment $s$ are always convex, but they may define invalid ears that cannot be used for triangulation.}
	\label{fig:all_ears}
\end{figure*}

\section{Background: classical earcut}
\label{sec:earcut_quad}
In this section we introduce the basics of the earcut algorithm, also fixing the notation.
Given a simple polygon $P$ defined by the cyclic list of its vertices $\{v_1,v_2,\dots,v_{n}\}$, a vertex $v \in P$ is \emph{convex} if its internal angle is less than $\pi$, and is \emph{concave} otherwise. For two non consecutive points $v_i,v_j \in P$, the segment $\overline{v_iv_j}$ is a \emph{diagonal} of $P$ if it is completely contained in the polygon. Given a convex vertex, if its two adjacent vertices form a diagonal, then this detaches a triangle -- or \emph{ear} -- from $P$.  It is known that any simple polygon contains at least two ears~\cite{meisters1975polygons}; the algorithm that progressively detaches all of them to construct a triangulation is called earcut, and is arguably the simplest triangulation algorithm known~\cite{eberly2008triangulation}. 

Despite its simplicity, earcut is fairly inefficient. A naive implementation has complexity $O(n^3)$, with $n$ being the number of polygon vertices. Precomputing separate lists for convex and concave vertices reduces complexity to $O(n^2)$~\cite{eberly2008triangulation}, still far from the best known triangulation algorithm for simple polygons, which promise linear time complexity (Section~\ref{sec:related}). Nevertheless, when it comes to actual coding, earcut is always a tempting solution due to its ease of implementation.

What makes earcut inefficient is the diagonal test. For any candidate ear centered at a convex vertex the algorithm must verify if its left and right neighbors form a diagonal. This amounts to ensure that the triangle described by these three vertices does not contain any other polygon vertex, which can be done in linear time by testing them all. Since the triangulation of a polygon with $n$ vertices contains $n-2$ triangles, and testing an ear is linear, the overall cost is quadratic at best.

\begin{algorithm}
	\Indm\Indmm
	\SetKwData{Left}{left}
	\SetKwData{Up}{up}
	\SetKwInOut{Input}{input}
	\SetKwInOut{Output}{output}
	\Indp\Indpp
	\Input{a simple polygon $P = \{v_1,v_2,\dots,v_{n}\}$, with vertices sorted so that $(v_1,v_n)$ are the endpoints of the constrained segment}
	\Output{a triangulation of $P$}
	\BlankLine
	\textcolor{gray}{// use a doubly linked list for $P$. Cost of update is $O(1)$}\\
	$P = \{  n, 1, 2, \dots, n-1 \}$ \textcolor{gray}{// prev}\\
	$N = \{  2, 3, \dots, n, 1 \}$ \textcolor{gray}{// next}\\
	
	\BlankLine
	\textcolor{gray}{// pre-compute internal ears. Cost is $O(n)$}\\
	$E = \emptyset$\\
	\For{$i= 2,3,\dots, n-1$}
	{
		\If{$v_i$ is a convex vertex}
		{
			append $v_i$ into $E$
		}
	}
	
	\BlankLine
	\textcolor{gray}{// process internal ears. Cost is $O(n)$}\\
	\While{$\vert E \vert > 0$}
	{
		$v$ = extract one ear from $E$\\
		make triangle $P(v), v, N(v)$\\
		
		\vspace{1em}
		\textcolor{gray}{// update adjacencies. Cost is $O(1)$}\\
		$N(P(v)) = N(v)$\\
		$P(N(v)) = P(v)$\\
		
		\vspace{1em}
		\textcolor{gray}{// check if prev or next are new ears. Cost is $O(1)$}\\
		\If{$P(v) \notin E \cup \{ v_1,v_n \}$ and $P(v)$ is convex}
		{
			append $P(v)$ into $E$\\
		}
		\If{$N(v) \notin E \cup \{ v_1,v_n \}$ and $N(v)$ is convex}
		{
			append $N(v)$ into $E$\\
		}
	}
	
	\caption{Linear Earcut}
	\label{alg:linear_earcut}
\end{algorithm}

\section{Linear earcut}
\label{sec:earcut_linear}
In this section we introduce our simplified version the earcut algorithm, also discussing its complexity. The proof of correctness and convergence of the algorith will be given in Section~\ref{sec:correctness}.

In Algorithm~\ref{alg:linear_earcut} we show a pseudo code implementation of our linearized earcut. Our code is based upon the efficient implementation described in~\cite{eberly2008triangulation}, which is further simplified to fully exploit the special nature of our polygons. Given an ordered chain of vertices $\{v_1,v_2,\dots,v_{n}\}$ describing a simple polygon, and assuming that $\overline{v_1,v_n}$ is the constrained segment we want to insert in the mesh, the algorithm proceeds as follows: we first initialize a doubly linked list representation of the polygon, which amounts to two vectors of length $n$ encoding, for each point, its previous and next vertices along the chain. This representation is extremely efficient, as deleting a node from the polygon amounts to updating the prev and next information from its neighbors, excluding it from the chain (lines 15-17 in the pseudo code). We then process all the vertices but the extrema of the constrained segment, and check whether they are convex or concave. All convexity checks are performed with exact orientation predicates~\cite{shewchuk1997adaptive}, hence the algorithm is numerically robust.  Differently from standard earcut, convex vertices are directly deemed as valid ears, as they do not necessitate the diagonal test (a formal proof is given in Section~\ref{sec:int_ears}).  Finally, we cut all ears: for each ear centered at a convex vertex $v$, we first create a triangle with the previous and subsequent vertices in the chain (lines 13-14 in the pseudo code), and then remove $v$ from the polygon. Finally, if the extrema of the ear were not convex vertices, we check whether they have now become convex, and if so we append them to the ear list. The algorithm terminates when all ears have been cut, resulting in a triangulation of the input polygon.



\subsection{Complexity}
\label{sec:complexity}
It is easy to verify that the aforementioned algorithm runs in deterministic linear time, which means that its complexity is liner in the number of polyon vertices in the worst case scenario. The pre-computation of internal ears (lines 4-10) amounts to compute $n-2$ internal angles (the extrema of the constrained segment are not considered), and is therefore $O(n)$. The while loop (lines 11-25) is executed as many times as the number of ears in the polygon. We know from Euler that a simple polygon with $n$ vertices can be triangulated exactly with $n-2$ elements, which means that the code in the loop will be executed exactly $n-2$ times. Inside the loop, we have the generation of the triangle, which is $O(1)$, the update of the doubly linked list, which is $O(1)$, and the check for new ears, which is restricted only to the sides of the ear we just cut, and is therefore $O(1)$ too. Therefore, the whole complexity is $\Theta(n)$.

\section{Proof of correctness}
\label{sec:correctness}
We demonstrate that our linearized earcut algorithm is guaranteed to converge to a valid triangulation. The outline of the proof is as follows:
we first characterize the class of polygons under which our method is guaranteed to work (Section~\ref{sec:polygon}). Then, we prove that all \emph{internal} ears (i.e. all convex vertices but the extrema of the constrained segment) can be safely cut in $O(1)$ sidestepping the diagonal test (Section~\ref{sec:int_ears}). Conversely, the two \emph{lateral} ears (i.e. the extrema of the constrained segment) may not be valid, and always require a diagonal test before cutting (Section~\ref{sec:lat_ears}). Finally, we prove that for any polygon in our class of study there always exist an internal ear, thus guaranteeing linear time convergence in the worst case scenario (Section~\ref{sec:conv}).

\begin{figure}[h!t]
	\centering
	\includegraphics[width=.9\columnwidth]{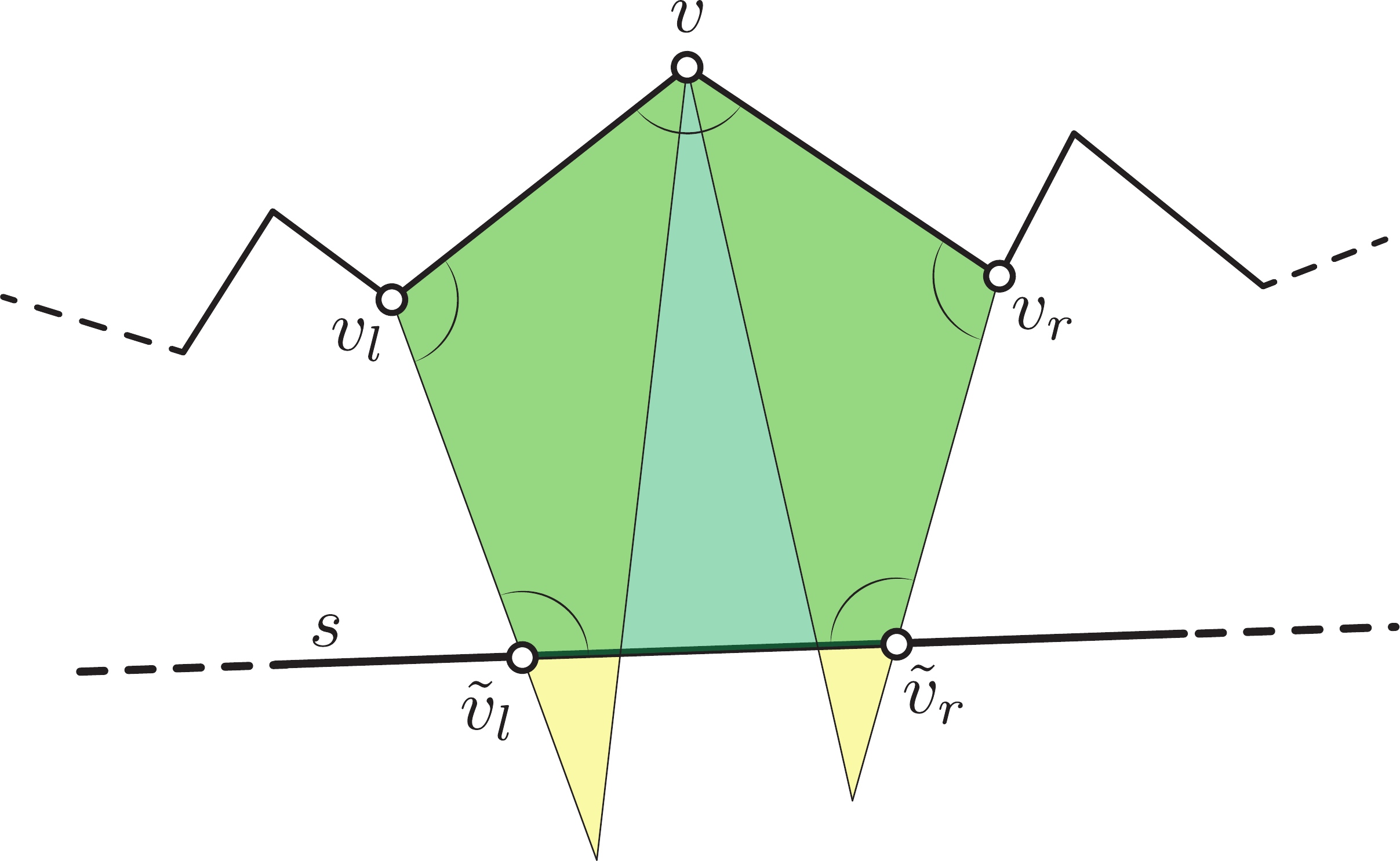}
	\caption{Any convex vertex $v$ along the chain that indirectly connects the endpoints of the constrained segment $s$ is guaranteed not to contain any other polygon vertex inside its ear. This can be proved by observing that $v$ is also a vertex of a convex sub-polygon (green shaded area). Convexity guarantees that all non subsequent vertices of the green polygon form valid ears. therefore, $v$ is also a valid ear for the original polygon.}	
	\label{fig:good}
\end{figure}

\subsection{Polygon properties}
\label{sec:polygon}
We are interested in tessellating polygons that arise in the context of constrained triangulations, when a new segment $s$ is to be inserted in a previously existing triangle mesh $M(V,T)$. The outer perimeter of all triangles in $T$ intersected by $s$ defines a polygon having the segment as diagonal. Halving such polygon along $s$ defines two sub-polygons, which must be triangulated in order to transform the segment $s$ into an edge of the mesh $M$ (Figure~\ref{fig:pocket}).

\begin{figure*}
\centering
\includegraphics[width=\linewidth]{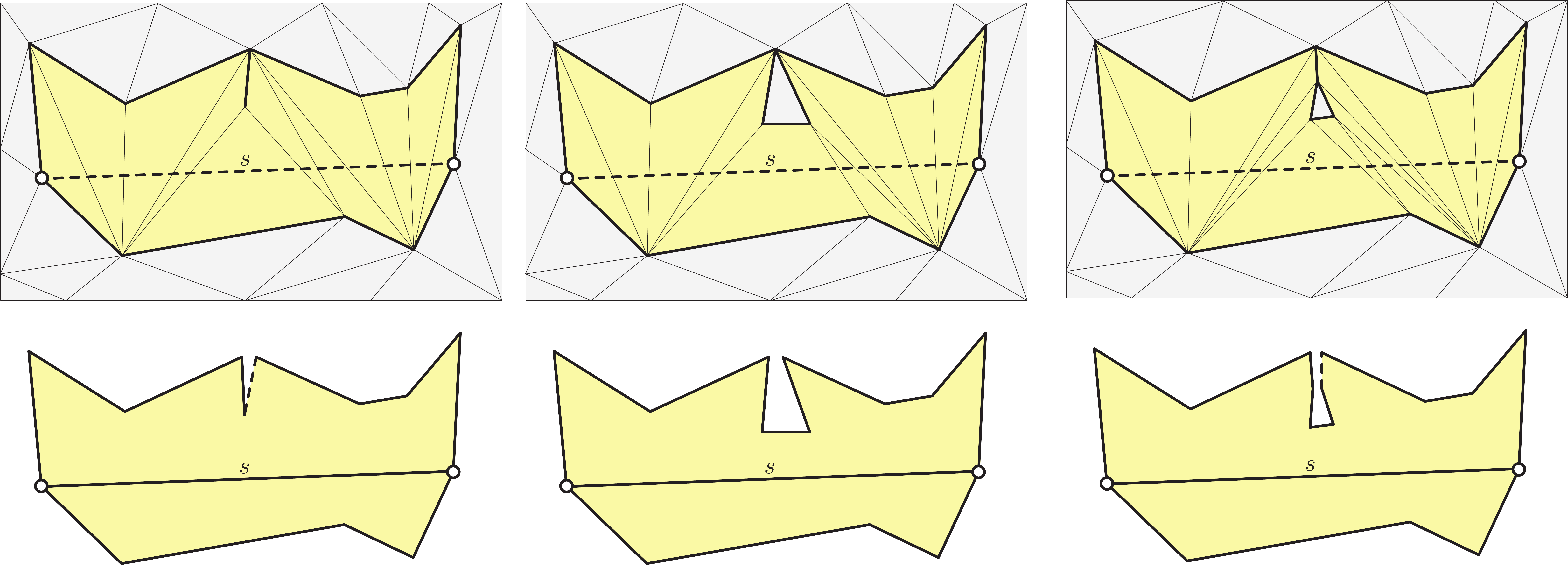}
\caption{Top: the outer profile of all triangles intersected by a constrained segment may enclose edges of the underlying mesh that are not intersected by $s$, generating dangling edges (left), holes (middle) or a combination of both (right). Extracting the polygons with a topological approach that marches on the underlying mesh allows to correctly handle these pathological cases, duplicating vertices with more than two neighbors along the chain (bottom). As a result, from a topological point of view the polygons are always simple, though in these pathological cases they will contain geometrically coincident vertices or edges.}
\label{fig:corner_cases}
\end{figure*}

As observed in~\cite{de1992line} polygons that arise in this context are simple, meaning that they do not self intersect and do not contain internal holes. Note that edges of the original mesh that do not intersect the segment may remain trapped inside the polygon, generating hanging edges, holes, and a combination of both. Nevertheless, extracting the polygon border with a method that marches on the underlying mesh, such as~\cite{shewchuk2015fast}, guarantees that non simple vertices are duplicated, always leading to a topologically simple polygon which may occasionally contain geometrically coincident vertices if these pathological cases occur (Figure~\ref{fig:corner_cases}).

They key observation we make in this paper is that these polygons belong to a restricted class of shapes which makes them much easier to triangulate than general simple polygons. In fact, given a segment $s$ and the two polygons that base upon it, we observe that despite concave, a portion of $s$ must be visible from any point inside the polygons that have it as a base. This observation was already made in~\cite{de1992line}, and stems from the fact that polygons we wish to triangulate are made of portions of triangles that are intersected from the new segment, hence the convexity of each sub-element guarantees visibility (Figure~\ref{fig:pocket}). More formally, polygons of this kind are referred to as \emph{weakly visible}\cite{avis1981optimal}. This property avoids the presence of severe (curl-like) concavities, permitting us to sidestep the diagonal test in our modified earcut.



\subsection{Internal ears}
\label{sec:int_ears}
Given a polygon $P$, we demonstrate that any convex vertex $v$ along the chain that indirectly connects the endpoints of the constrained segment $s$ forms a valid ear. In other words, denoting with $v_l,v_r$ the two vertices at the immediate left and right of $v$, we prove that the triangle $\widehat{v_l v v_r}$ does not contain any other vertex of $P$ in its interior. We prove our thesis by showing that $v_l, v, v_r$ are also vertices of a convex sub-polygon $\Omega \subseteq P$. Convexity guarantees that any pair of non adjacent vertices in $\Omega$ forms a valid diagonal, included the one connecting $v_l$ and $v_r$.

Without loss of generality, let us focus on the left side of $v$. By symmetry, the same construction can be generated at its right side. If $v_l$ is not an endpoint of $s$, then the edge $\overline{v_l v}$ belongs to a triangle $t_l$ in the underlying mesh, and this triangle has its third vertex at the opposite side of $s$. This must always be the case, because if such point was on the same side of $s$, then $t_l $ would not intersect the segment in the first place, and would not be part of polygon $P$. The edge connecting the triangle vertex opposite to $\overline{v_l v}$ with $v_l$, intersects segment $s$ at a point $\tilde{v}_l$. Similarly, there exists a twin vertex $\tilde{v}_r$, obtained replicating the same construction at the right side of $v$.

Points $ \{  v, v_l, \tilde{v}_l,\tilde{v}_r, v_r \}$ form a pentagon $\Omega \subset P$ (Figure~\ref{fig:good}). It can be easily shown that $\Omega$ is provably strictly convex, in fact:



\begin{itemize}
	\item by our initial hypothesis, $v$ is a convex vertex of $P$, hence its angle is strictly less than $\pi$ also in $\Omega$;
	\item $v_l$ is a vertex internal to the triangle $t_l$, therefore its angle is strictly less than $\pi$;
	\item $\tilde{v}_l$ is defined by the intersection of a triangle edge with segment $s$. This intersection partitions $2\pi$ into four angles, all strictly less than $\pi$;
	\item by symmetry, angle bounds for $v_l,\tilde{v}_l$ also hold for $v_r$ and $\tilde{v}_r$
\end{itemize}
Since $\Omega$ is convex, any triplet of consecutive vertices forms a valid ear, including the ear centered at $v$. 
\qed\\

Note that, in case either $v_l$ or $v_r$ are endpoints of $s$, then $\Omega$ is a quadrilateral. If both are endpoints of $s$, then $\Omega$ is a triangle and is coincident with $P$. In both cases, the inner angles of $\Omega$ are still strictly bounded by $\pi$, hence its convexity and the validity of all its ears are verified. In Figure~\ref{fig:all_ears} we show all convex sub-polygons that protect the internal ears for the example shown in Figure~\ref{fig:pocket}.

%

\subsection{Lateral ears}
\label{sec:lat_ears}
The extrema of the constrained segment $s$ always form convex vertices with respect to $P$. This can be easily verified by observing that if their angle was greater or equal to $\pi$, than the triangle that contains them would not intersect $s$ in the first place. However, the construction described in Section~\ref{sec:int_ears} does not apply to \emph{lateral} ears, because one of their sides coincides with $s$, and the existence of a convex sub-polygon that contains it and is fully contained in $P$ is not guaranteed. In the bottom right part of Figure~\ref{fig:all_ears} we show a failure example where the triangle span by a lateral ear is not even contained in $P$. Note that cases in which lateral ears form valid triangles may still occur, but they cannot be safely cut without a diagonal test, hence the computational cost of processing them is $O(n)$. For this reason, we never consider lateral ears in our triangulation algorithm.

\subsection{Existence of internal ears}
\label{sec:conv}
To prove convergence in deterministic linear time, we demonstrate that for any polygon defined in Section~\ref{sec:polygon} there always exist an internal ear which can be cut in $O(1)$ (Section~\ref{sec:int_ears}). As a starter, we observe that any simple polygon has at least three convex vertices. This can be proved by observing that the sum of internal angles of a simple polygon with $n$ vertices is always $(n-2)\pi$. Let us assume that there are only two convex vertices, with angles $\alpha,\beta>0$. By the definition of concave vertex, the sum of the angles of the remaining $n-2$ vertices must be equal or greater than $(n-2)\pi$. Summing up the angles of the two convex vertices we obtain $(n-2)\pi + \alpha + \beta$, which is already greater than the overall sum of all internal angles of a polygon, $(n-2)\pi$, leading to a contradiction. But, if a polygon contains at least three convex vertices and exactly two lateral ears, the third convex vertex must be an internal ear. \qed

\begin{figure}
	\includegraphics[width=\linewidth]{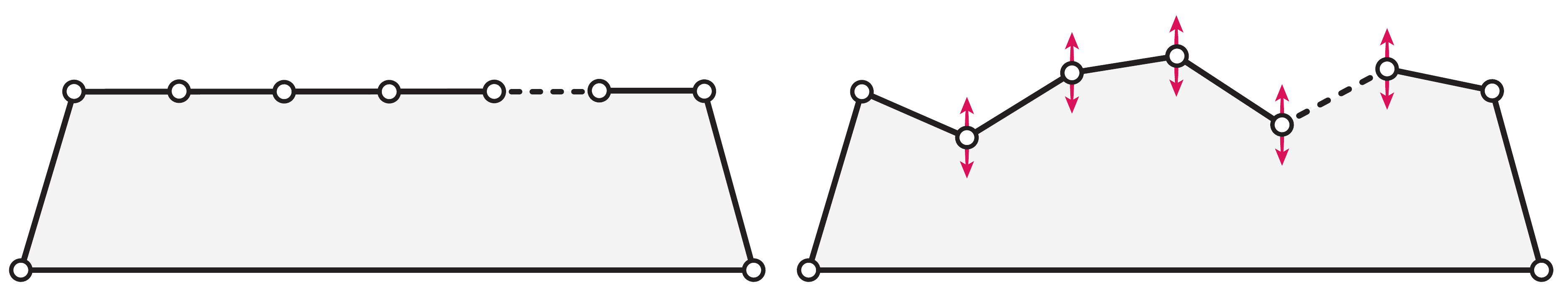}
	\caption{Parametric test polygons used for the comparison with~\cite{shewchuk2015fast}. The lower edge corresponds to the constrained segment, the upper side can accommodate a varying number of vertices (all collinear in the left model, randomly displaced along the vertical axis in the right model).}
	\label{fig:test_polys}
\end{figure}

\begin{figure}[t]
	\includegraphics[width=\linewidth]{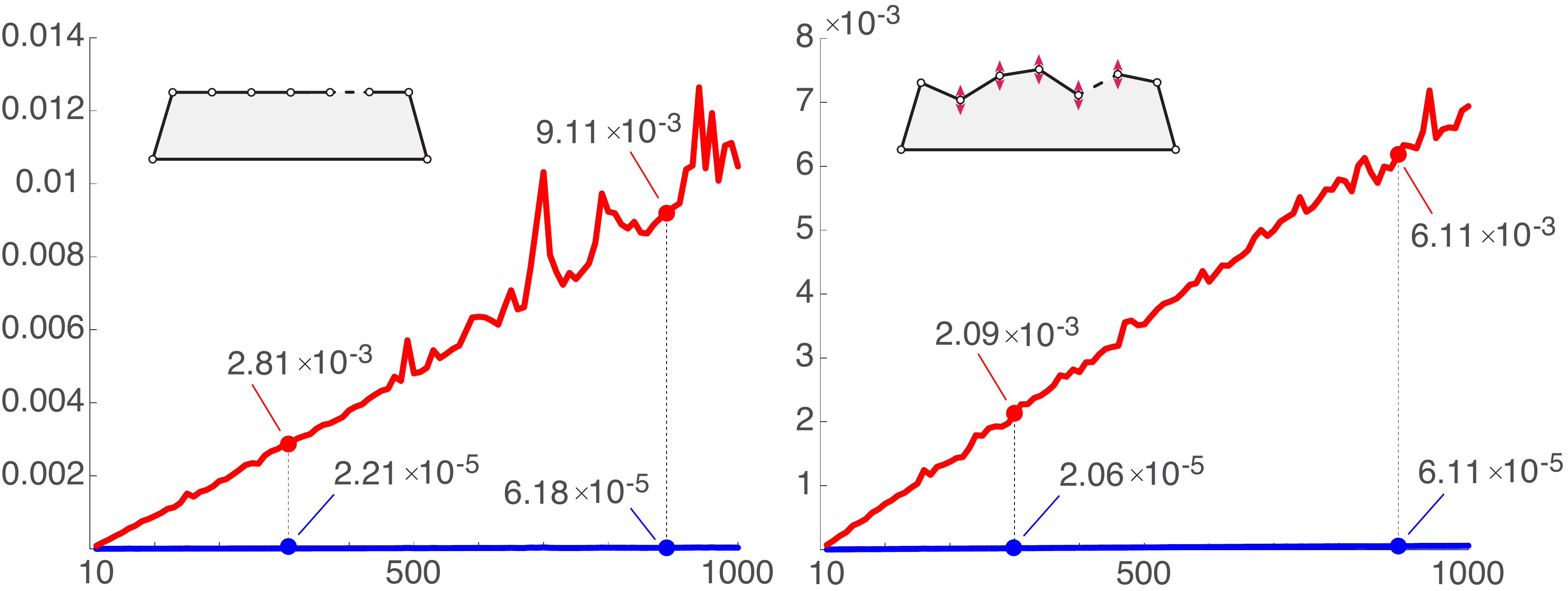}
	\caption{Time comparison between our method (blue) and the randomized algorithm proposed in~\cite{shewchuk2015fast} (red), obtained considering the two polygons in Figure~\ref{fig:test_polys}. Horizontal axis reports number of polygon vertices, vertical axis running times (in seconds). For each polygon size, we considered the average running time over 1000 attempts. Both algorithms exhibit linear growing times w.r.t. to polygon size, but our method is approximately two orders of magnitude faster.}
	\label{fig:comp}
\end{figure}

\section{Experimental validation}
\label{sec:results}

We have implemented our linearized earcut algorithm and compared it against the most recent prior art, which consists in the algorithm proposed by Shewchuk and Brown in~\cite{shewchuk2015fast}. Just as our algorithm, their method is specialized to triangulate the class of polygons described in Section~\ref{sec:polygon}, but it is randomized and hence has only \emph{expected} linear time complexity. Their algorithm produces a Constrained Delaunay Triangulation (CDT) of the input polygon, obtained using Chew's algorithm~\cite{chew1990building} to tessellate in expected linear time sub-polygons that violate the Delaunay criterion. We observe that a direct comparison would not be fair because our constrained triangulations do not necessarily have the Delaunay property. Therefore, we have repimplemented Shewchuk and Brown's algorithm while omitting Chew's module and the incircle tests, so that the algorithm produces only a general triangulation without the overhead required to obtain the Delaunay property. We used this version to conduct our experimental comparison.



Both algorithms were implemented in C++, using a MacBook Pro equipped with an Intel Core i5 2.9GHz and with 16GB of RAM as testing hardware. Considering the simplicity of our method, turning the pseudo code provided in Algorithm~\ref{alg:linear_earcut} into actual code took us less than one hour (a reference implementation can be found inside CinoLib~\cite{cinolib} at the following link  \url{https://github.com/mlivesu/cinolib/blob/master/include/cinolib/segment_insertion_linear_earcut.h}). Also~\cite{shewchuk2015fast} is considered relatively simple to code, and the authors reported five hours to implement the algorithm starting from their pseudo code. In our personal experience we needed two days of work to fully understand the algorithm and make a computer program out of it. Implementing Chew's submodule (which we omitted) might also require some extra time.

For the experiments, we followed the same validation scheme used in~\cite{shewchuk2015fast}, which measured growth of running times w.r.t. the input size, measuring it on two parametric polygons with growing number of vertices. Despite these two polygons do not exhaustively represent the class of cases that can arise in real applications, they are complex enough to reveal critical configurations and bottlenecks~\cite{shewchuk2015fast}. For completeness, we also performed a third experiment `\emph{in the wild}`, using the two triangulation algorithms inside the pipeline for the computation of mesh arrangements proposed in~\cite{cherchi2020}, and launching the software on the 4K intersecting meshes contained in the Thingi10K dataset~\cite{zhou2016thingi10k}.

\textbf{Parametric polygons.} We tested both algorithms on the parametric shapes depicted in Figure~\ref{fig:test_polys}, considering polygons having from 10 to 1000 vertices. For each polygon, we averaged running times across 1000 different runs, so as to void biases depending from external factors. Results are shown in Figure~\ref{fig:comp}. Both algorithms exhibit linear growth in the running times w.r.t. the input size, but our algorithm exhibits less fluctuations and is also sensibly faster. One of the possible reasons for this difference is that
~\cite{shewchuk2015fast} during its iterations may produce triangles that conflict with previously generated triangles, which must be removed. This not only introduces unnecessary delays in the algorithm, but also requires some sort of mesh data structure to handle the topological changes and inspect the neighborhood of each newly generated triangle in order to check whether a conflict exists or not. In contrast, earcut generates only legal triangles that will appear in the output tessellation, and does not require any supporting mesh data structure during its execution. Note that if a CDT is to be constructed, also our method would need at least the ability to find the vertices opposite to a given edge to perform incircle tests, as well as an edge flip operator to secure the Delaunay property. Even in that case, no extra cost will be paid to remove illegal triangles.

\textbf{Mesh Arrangements in the wild.} 
Computing a mesh arrangement consists in refining an input triangle soup in order to incorporate intersection points in the connectivity. The typical pipeline works by refining each triangle separately, adding intersection points first, and then including constrained segments that arise when two triangles intersect. This latter step can be accomplished by using constrained triangulation algorithms.
We run the pipeline proposed in~\cite{cherchi2020} twice, once using our linearized earcut method for segment insertion, and once using the method proposed in~\cite{shewchuk2015fast}. 
The time required for the segment insertion step, and the consequent triangulation of the polygons,  in all the 4408 intersecting meshes in Thingi10K was 18 minutes with our method, and 23 minutes with~\cite{shewchuk2015fast}. In most of the cases the differences between the two methods were negligible (i.e. less than 1e-5 seconds), but overall, our method was faster in 3969 models out of 4408, showing that even in real cases it can be consistently faster than prior art.
\section{Conclusions and future works}
\label{sec:conclusions}
We presented a novel algorithm to triangulate in deterministic linear time a restricted class of planar polygons that arise in the context of constrained triangulations. Tessellations of this kind often arise in scientific computing, hence the proposed method is of practical relevance. A few deterministic linear time methods were already known in literature, but they are all based on complex trapezoidation schemes, and are difficult to implement.
As a result, 
easier (though sub optimal) methods that run in quadratic, logarithmic, or non deterministic linear time at best are used in practice instead. 

Our method merges optimality guarantees with ease of implementation, because it is based on a further simplification of the earcut algorithm~\cite{eberly2008triangulation}, which is arguably one of the easiest (though fairly inefficient) triangulation algorithms to implement. We have shown that by omitting the diagonal test earcut achieves optimality, and can still guarantee the correct result if the input polygon is in the class of our interest. We also provided rigorous proof of our findings, and practical evidence that the proposed algorithm is indeed faster in practice. Due to these key features, we expect future codes for the generation of constrained triangulations to readily adopt our tool.


\subsection{Future works}
Our method is not concerned with mesh quality, hence triangles can be arbitrarily badly shaped. 
There are two major ways to improve triangulation quality: one is to construct a Delaunay triangulation using a method similar to the one used in~\cite{shewchuk2015fast}, which is suitable to our pipeline. Alternatively, one could use either a randomized or a prioritized version of our linearized earcut, which modifies the processing order of the ears and tends to produce much better triangulations than the standard version based on sequential processing~\cite{held2001fist}. Note that both Delaunay and prioritized earcut alter the complexity of the algorithm, whereas randomized earcut remains deterministic linear.
	
Besides quality, this work opens for two interesting lines of future works. On the one side, if similar properties could be proved also in 3D, this might lead to optimal time algorithms for constrained volumetric meshing. This extension is far from being obvious though. For the 2D case one of the key ingredients to prove the convexity of the sub-polygon enclosing a convex vertex was that given an edge, the triangle containing it had always its third vertex at the opposite side of the constrained segment. In 3D constraints can be both planar polygons or segments. The notion of \textit{being at the opposite side of a segment} is not well defined. Moreover, there can be tetrahedra that intersect the constraint and have two vertices on one side of it and the other two at the other side. It is not clear how this configuration can be handled in the current logical scheme of our proof. Nonetheless, in 3D not all concave polytopes can be triangulated without additional (Steiner) points~\cite{schonhardt28}, and even deciding whether this is possible is NP-Hard~\cite{ruppert92b}. Non decomposable polyhedra can arise at any step of the pipeline, causing a deadlock.  The second interesting line of research regards parallelization. In~\cite{eder2018parallelized} a parallel version of the standard earcut method was introduced. An adaptation of the same parallelization scheme to our linearized earcut seems possible, and might lead to sub linear segment insertion for constrained triangulation.



\ifCLASSOPTIONcompsoc
  \section*{Acknowledgment}

Gianmarco Cherchi gratefully acknowledges the support to his research by PON R\&I 2014-2020 AIM1895943-1.
Marco Livesu and Marco Attene's work was partly supported by EU ERC Advanced Grant CHANGE  No. 694515.

\ifCLASSOPTIONcaptionsoff
  \newpage
\fi


\bibliographystyle{IEEEtran}
\bibliography{00_main}


\begin{IEEEbiography}[{\includegraphics[width=1in,height=1.25in,clip,keepaspectratio]{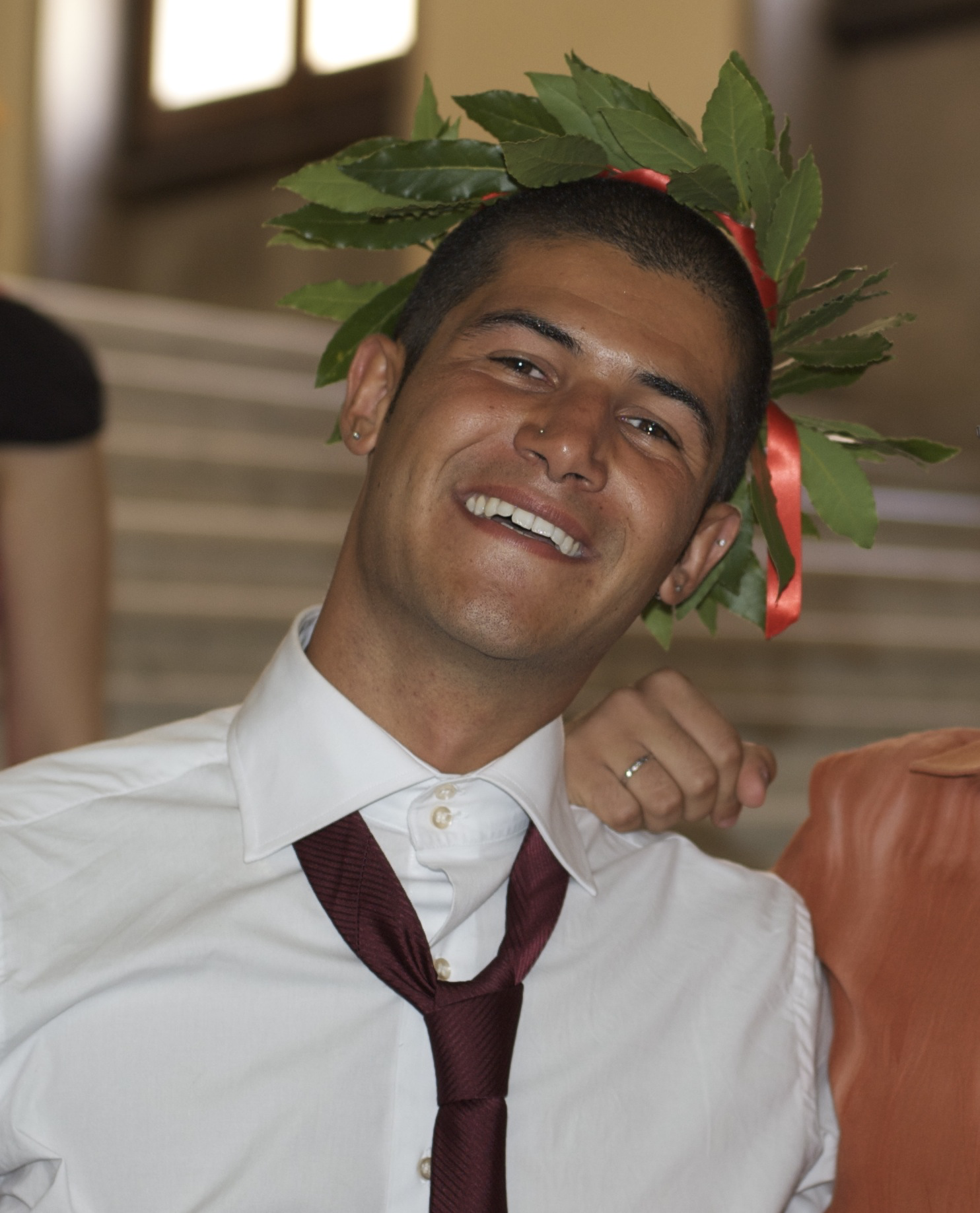}}]{Marco Livesu}
is a tenured researcher at the Institute for Applied Mathematics and Information Technologies of the National Research Council of Italy (CNR IMATI). He received his PhD at University of Cagliari in 2014, after which he was post doctoral researcher at the University of British Columbia, University of Cagliari and CNR IMATI. His main research interests are in computer graphics and geometry processing.
\end{IEEEbiography}

\begin{IEEEbiography}[{\includegraphics[width=1in,height=1.25in,clip,keepaspectratio]{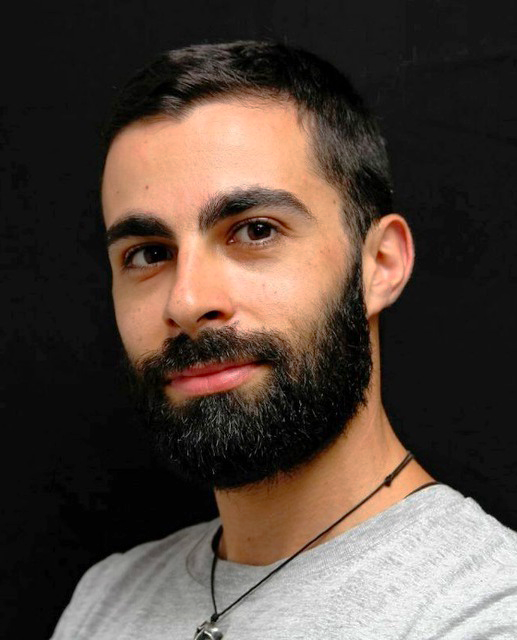}}]{Gianmarco Cherchi}
is a computer science researcher at the Department of Mathematics and Computer Science of the University of Cagliari (UNICA). He received his Ph.D. at the University of Cagliari in 2018. Since his first year of university, he is collaborating with the CG3HCI (Computer Graphics, Computation Geometry and Human-Computer Interaction – Cagliari Group) Research group. His main interests are in computer graphics and geometry processing.
\end{IEEEbiography}

\begin{IEEEbiography}[{\includegraphics[width=1in,height=1.25in,clip,keepaspectratio]{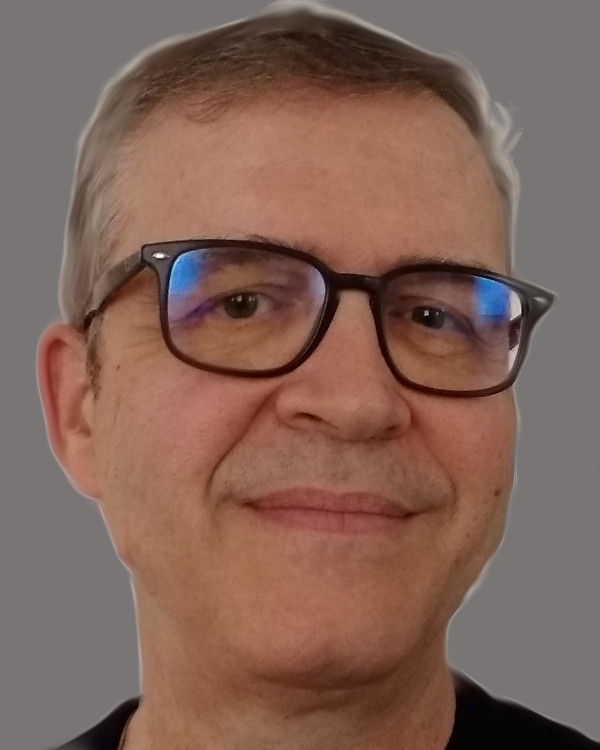}}]{Riccardo Scateni}
is professor of Computer Science at the Department of Mathematics and Computer Science of the University of Cagliari since 2001 where later he founded the CG3HCI (Computer Graphics, Computation Geometry and Human-Computer Interaction – Cagliari Group). Before joining UniCA he was post-doc at IBM Kingston, USA, at CERFACS, Toulose and researcher and senior resarcher at CRS4, Cagliari, Italy. His current research interests are in geometry processing and additive fabrication.
\end{IEEEbiography}

\begin{IEEEbiography}[{\includegraphics[width=1in,height=1.25in,clip,keepaspectratio]{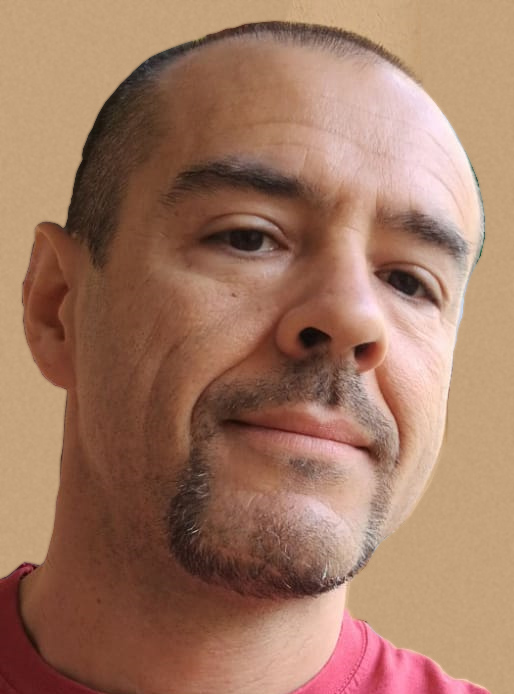}}]{Marco Attene}
is a Senior Researcher at the Institute of Applied Mathematics and Information Technologies (IMATI) of the italian CNR, where he contributes to advance geometry processing with particular focus on mesh repairing and 3D printing applications. Marco has been the promoter of key collaborations and joint research programs between IMATI and universities in Europe, USA, Asia and New Zealand. He manages several open source software projects, and his “MeshFix” system received the SGP Software Award in 2014. He has been program chair of international conferences and is an associate editor of international journals in the area and, since 2019, he is the general chair of the Graphics Replicability Stamp Initiative.
\end{IEEEbiography}



\end{document}